\documentclass[12pt]{article}
\usepackage{epsfig}
\usepackage{citesort}
\usepackage{amssymb,amsfonts}
\newcommand{\beq}{\begin{equation}}
\newcommand{\eeq}{\end{equation}}
\newcommand{\bea}{\begin{eqnarray}}
\newcommand{\eea}{\end{eqnarray}}
\newcommand{\pslash}{p\hspace{-.18cm}/}

\begin{document}
\title{\bf Thermodynamics of two-colour QCD and the Nambu Jona-Lasinio model
\footnote{Work supported in part by INFN and BMBF}}
\author{
\bf Claudia Ratti $^a$ and Wolfram Weise $^{a, b}$\\ 
\\
$^a$ {\small Physik-Department, Technische Universit\"at M\"unchen, D-85747 
Garching, Germany} \\
$^b$ {\small 
ECT*, I-38050 Villazzano (Trento), Italy}
      }
\date{\today}
\maketitle
\begin{abstract}
We investigate two-flavour and two-colour QCD at finite temperature 
and chemical potential in comparison with a corresponding Nambu and 
Jona-Lasinio model. By minimizing the thermodynamic potential of the system, 
we confirm that a second order phase
transition occurs at a value of the chemical potential equal to half the mass
of the chiral Goldstone mode. For chemical potentials beyond this value the scalar
diquarks undergo Bose condensation and the diquark condensate is nonzero. We 
evaluate
the behaviour of the chiral condensate, the diquark condensate, the baryon
charge density and the masses of scalar diquark, antidiquark and pion,
as functions of the chemical potential. Very good 
agreement is found with lattice QCD ($N_c=2$) results. We also compare with a 
model based on leading-order chiral effective field theory.
\end{abstract}
\vfill\eject
\section{Introduction}
The phase structure of QCD has been subject of intense investigations in recent
years. Precise numerical data have become available concerning
QCD thermodynamics at high temperature via large-scale computer simulations on 
the lattice
(for a review see~\cite{Karsch:2001cy}). The study of full QCD at 
finite baryon density is still a formidable challenge, due to the limitations 
of standard Monte Carlo simulations when applied to systems at finite chemical
potential (for recent results see~\cite{Muroya:2003qs,Karsch:2003zq}). 
Present developments are aimed at improved strategies~\cite{Fodor:2002} to deal
with the fact that the determinant of the Euclidean Dirac operator becomes
complex at finite chemical potential. 

An interesting perspective of finite-density QCD is the emergence of colour 
superconductivity (CSC). This was revealed first by calculations based on
one-gluon exchange: Barrois, Bailin and Love 
\cite{Barrois:1977xd,Bailin:1984bm} and 
later Iwasaki and Iwado~\cite{Iwasaki:1995ij} pointed out
that the induced attractive force near the Fermi surface creates quark Cooper 
pairs resulting 
in CSC in the case of QCD at low temperature and high density.
In the late nineties, using an instanton model of the effective interaction, 
Alford, Rajagopal and Wilczek~\cite{Alford:1998zt,Rajagopal:1999cp} and Rapp, 
Sch\"afer, Shuryak and Velkovsky~\cite{Rapp:1998zu} argued that the energy gap 
is expected to be of the order of 100 MeV. 

No first principle computations exist at this moment concerning
the phenomenon of colour superconductivity in full $N_c=3$ QCD. One 
response to this situation has been to start from simpler QCD-like theories 
with additional antiunitary 
symmetries that guarantee the Fermion determinant to be real at non-zero 
chemical potential and therefore
allow the study of such theories on the lattice. Examples of such explorations
include QCD with two colours and fundamental quarks and 
QCD with an arbitrary number of colours and adjoint quarks~\cite{Kogut:2000ek}.
The physics of both 
these theories is quite
different from full three-colour QCD. Nevertheless these differences are easily
understood and classified. Knowledge of the critical conditions for phase 
transitions in 
these schematic cases may offer qualitative clues about critical phenomena
encountered in three colour QCD, such as diquark condensation. 

In two-colour QCD, diquarks can form colour singlets which are
the baryons of the theory. The lightest baryons and the lightest 
quark-antiquark excitations
(pions) have a common 
mass, $m_{\pi}$, and this spectrum determines the properties of the ground
state for small chemical potentials. General arguments~\cite{Halasz:1998qr} 
predict a phase 
transition from the vacuum to a state with finite baryon density at a critical
chemical potential $\mu_c$, which is the lowest energy per quark
that can be realized by an excited state of the system.  
This state is populated by light diquarks, and one expects $\mu_c=m_{\pi}/2$.
The Bose-Einstein condensation of diquarks, with
nonzero baryon number, can be interpreted as baryon charge superconductivity.

The $(T,\mu)$ phase diagram of QCD with two colours has been studied
by Dagotto {\it et al.} using a mean-field model of the lattice action
\cite{Dagotto:1986gw,Dagotto:1987xt}. 
The smallness of $\mu_c$ has been
exploited to study the zero temperature phase transition using a chiral
effective Lagrangian extended to the flavour symmetry $SU(2N_f)$
\cite{Kogut:2000ek,Kogut:1999iv,Splittorff:2000mm,Splittorff:2001fy,Splittorff:2002xn}. Other 
approaches to two-colour QCD have also 
been explored, based for example on a random matrix 
model~\cite{Vanderheyden:2001gx,Klein:2004hv} and on the 
renormalization group~\cite{Wirstam:2002be}. Several of these model 
calculations have been verified by lattice simulations
\cite{Lombardo:1999gr,Lombardo:2003uu,Hands:1999md,Hands:2000ei,Hands:2000hi,Hands:2001ee,Aloisio:2000nr,Aloisio:2000if,Aloisio:2000rb,Liu:2000in,Muroya:2000qp,Muroya:2002ry,Alles:2000qi,Alles:2002st,Bittner:2000rf,Kogut:2001if,Kogut:2001na,Kogut:2003ju,Kogut:2002cm,Skullerud:2003yc}.

In the present paper we investigate the relationship between $N_c=2$ QCD and a 
corresponding Nambu and Jona-Lasinio (NJL) model
\cite{Nambu:1961tp,Nambu:1961fr,Vogl:1991qt,Klevansky:1992qe,Hatsuda:1994pi} 
in which gluonic degrees of freedom are ``integrated out'' and replaced by
a local four-point interaction of quark colour currents. This amounts to
effectively replacing the local colour gauge symmetry by a global one, 
with the assumption that coloured (gluonic) excitations are far removed from 
the low-energy spectrum and hence ``frozen''.
Similar models have already been used to 
study the QCD colour superconductivity phase with two
\cite{Berges:1998rc,Langfeld:1998yf,Buballa:2001wh,Buballa:2002wy,Blaschke:2003cv} and three
flavours~\cite{Nebauer:2001rb,Buballa:2001gj,Neumann:2002jm} (for a recent
review see~\cite{Buballa:2003qv}).
The specific aim of this work is to test the effectiveness of the NJL model,
with its dynamically generated quasiparticles, in reproducing the 
thermodynamics of two-colour QCD, and to compare
our results quantitatively with those obtained from recent lattice 
computations.
We study the behaviour of the chiral and diquark condensates, and of the
baryon density, as functions of temperature and chemical potential, both in 
the chiral limit and for finite values of the current quark masses. We 
investigate, again for both zero and finite quark masses, the two-colour QCD 
phase diagram in the $T$-$\mu$ plane. As further applications we evaluate 
the pion, diquark and antidiquark masses, as functions of the chemical 
potential.
We compare our results to lattice data and also to the predictions
from chiral effective field theory.
\section{Two colour NJL model}

Consider as a starting point the Lagrangian
\beq
\mathcal{L}=\bar{\psi}\left(x\right)\left(i\gamma^{\mu}\partial_{\mu}-\bf{m_0}
\right)\psi\left(x\right)-G_c\sum_{a=1}^{3}J^{a}_{\mu}\left(x\right)
J_{a}^{\mu}\left(x\right),
\label{lc}
\eeq
with a four point interaction that represents the local coupling between colour
currents $J^{a}_{\mu}=\bar{\psi}\gamma_{\mu}t^a\psi$ involving the quark
fields $\psi$ and the $SU(2)_{colour}$ generators $t_a$ with
$tr(t^at^b)=2\delta_{ab}$. Here $G_c$ is an effective coupling strength of
dimension (length$)^2$ and {\bf m$_0$} is the diagonal current
quark mass matrix.

In this paper we restrict ourselves to the case of two quark flavours 
($N_f=2$). In this case there are only two order parameters, the quark 
condensate 
$\langle\bar{\psi}\psi\rangle$ and the scalar diquark 
condensate, symbolically denoted by $\langle\psi\psi\rangle$. It is convenient 
to rewrite the 
interaction between quarks, by Fierz transformation, in terms of the colour
singlet pseudoscalar/scalar
quark-antiquark and scalar diquark channels. The resulting Lagrangian reads 
\bea
\mathcal{L}_{NJL}&=&\bar{\psi}\left(i\gamma^{\mu}\partial_{\mu}-{\bf m_0}\right)\psi
+\mathcal{L}_{q\bar q}+\mathcal{L}_{qq}+(\mathrm{colour~triplet~terms}),
\label{lag}
\\
\mathcal{L}_{q\bar q}&=&\frac G2\left[\left(\bar{\psi}\psi\right)^2+
\left(\bar{\psi}i\gamma_5\vec{\tau}\psi\right)^2\right],
\nonumber\\
\mathcal{L}_{qq}&=&\frac H2\left(\bar{\psi}i\gamma_5\tau_2t_2
C\bar{\psi}^T\right)\left(\psi^TCi\gamma_5\tau_2t_2\psi\right)
\nonumber
\eea
where $G$ and $H$ are 
constants which describe quark-antiquark and quark-quark interactions, 
respectively, $t_a$ are Pauli matrices in colour space and $\tau_i$ 
are Pauli matrices in flavour (isospin) space. We have introduced the 
charge conjugation operator for fermions:
\beq
C=i\gamma_0\gamma_2.
\eeq

The coupling constants $G$ and $H$ in the Lagrangian~(\ref{lag}) are fixed 
by Fierz transforming the colour-current interaction in~(\ref{lc}) to obtain
\beq
G=H=\frac 32G_c
\label{coeff}
\eeq
(see the Appendix for details).

As mentioned, the local $SU(N_c=2)$ gauge symmetry is replaced by global $SU(2)_{colour}$ 
in this model.
In the chiral limit, the Lagrangian~(\ref{lag}) is invariant under an 
enlarged flavour symmetry 
$SU(N_f)\times SU(N_f)\times U(1)\rightarrow SU(2N_f)$, 
which connects quarks and antiquarks: the so-called Pauli-G\"ursey 
symmetry, a characteristic feature of two-colour QCD. This symmetry relates 
pions and 
scalar diquarks. It is a natural ingredient of the ``equivalent'' NJL model, 
with eq.~(\ref{coeff}) relating the coupling constants of the
model Lagrangian.

Starting from the Lagrangian (\ref{lag}) and using standard bosonization
techniques, we introduce the auxiliary scalar ($\sigma$), pseudoscalar
triplet\footnote{Isovectors such as the pion field are denoted 
by $\vec{\pi}$.} ($\vec{\pi}$) and diquark ($\Delta$, $\Delta^*$) fields, 
thus obtaining
the following equivalent Lagrangian in the colour singlet sector:
\bea
\tilde{\mathcal{L}}&=&\bar{\psi}\left(i\gamma^{\mu}\partial_{\mu}-{\bf m_0}+
\sigma+i\gamma_5\vec{\tau}\cdot\vec{\pi}\right)\psi-\frac 12 \Delta^*\psi^T C
\gamma_5\tau_2t_2\psi
\nonumber\\
&+&\frac 12\Delta \bar{\psi}\gamma_5\tau_2t_2
C\bar{\psi}^T-\frac{\sigma^2+\vec{\pi}^2}{2G}-\frac{\left|\Delta\right|^2}{2H}.
\label{bosonization}
\eea 
It is useful to represent the quark fields by a bispinor defined
in the following way:
\beq
q\left(x\right)=\frac{1}{\sqrt{2}}\left({{\begin{array}{c} \psi(x)\\C\bar
{\psi}^T(x)\end{array}}}
\right).
\eeq
Furthermore, we introduce the matrix propagator 
\beq
S^{-1}\left(p\right)=\left({{\begin{array}{cc} {\pslash} -\hat{M} & \Delta
\gamma_5\tau_2t_2\\-\Delta^*\gamma_5\tau_2t_2 & {\pslash}-\hat{M}
\end{array}}}\right)
\eeq
(the inverse of the so-called Nambu-Gorkov propagator)
where we have defined
\beq
\hat{M}=(m_0-\sigma){\bf 1}-i\gamma_5\vec{\tau}\cdot\vec{\pi};
\eeq
here ${\bf 1}={\bf 1}_c\cdot{\bf 1}_f\cdot{\bf 1}_D$ is the unit matrix in 
colour, flavour and Dirac indices.
We consider the flavour-symmetric case with $m_u=m_d\equiv m_0$.
Integrating over $q(x)$ and $\bar q(x)$ we obtain the effective Lagrangian
in terms of the auxiliary field variables $\sigma, \vec{\pi}$, $\Delta$ and 
$\Delta^*$. 
It can be written as:
\beq
\mathcal{L}_{eff}=-\frac {\sigma^2+\vec{\pi}^2}{2G}-\frac{\left|\Delta
\right|^2}{2H}-i\int\frac{d^4p}{\left(2\pi^4\right)}\frac 12 \mathrm{Tr}\ln
\left(S^{-1}\left(p\right)\right).
\label{leff}
\eeq
The trace in this expression is taken over flavour, colour and Dirac 
indices, and the factor $\frac 12$ compensates for double-counting in the $q$ 
and ${\bar q}$ fields.

Solving the field equations
for $\sigma$, $\vec {\pi}$, $\Delta$ and $\Delta^*$ and working in the mean
field approximation\footnote{In the mean field approximation the fields are 
replaced by their expectation values for which we will later on continue using
the notation $\sigma$ and $\Delta$ for simplicity and convenience.}, we can 
evaluate their vacuum expectation values. The mean field value 
$\langle\vec{\pi}\rangle$ of the pseudoscalar isotriplet field is always equal 
to zero. The $\sigma$ field has a non-vanishing vacuum expectation value as a 
consequence of spontaneous chiral symmetry breaking, while the diquark fields
$\Delta$ and $\Delta^*$ are expected to have nonzero mean values only in
dense matter. An interesting limiting situation is encountered when $m_0=0$
(chiral limit) together with $\mu=0$. In this limit the extended $SU(2N_f)$
symmetry with $N_f=2$ (and $G=H$) implies that the thermodynamic potential
depends only on $R^2=\sigma^2+|\Delta|^2$ so that there is a degeneracy along
the circle with constant radius $R$. This case will be further discussed
in section~\ref{chiral}.

After solving the field equation for $\sigma$, we can work in terms of the 
effective quark mass $m$ which is related to $\langle\sigma\rangle$ through the
self-consistent gap equation
\beq
m=m_0-\langle\sigma\rangle=m_0-G\langle\bar{\psi}\psi\rangle.
\label{mass}
\eeq
Note that $\langle\sigma\rangle=G\langle\bar{\psi}\psi\rangle$ is negative in 
our representation, and $\langle\bar{\psi}\psi\rangle=\langle\bar{\psi}_u
\psi_u+\bar{\psi}_d\psi_d\rangle$ with $\langle\bar{\psi}_u
\psi_u\rangle=\langle\bar{\psi}_d\psi_d\rangle$.
\section{Parameter fixing}
The three parameters of the model are the ``bare'' quark mass $m_0$, 
a loop-momentum cutoff $\Lambda$ and the coupling strength $G=H$.
Even if we are considering the $N_c=2$ NJL model, we choose to reproduce the
known chiral physics in the hadronic sector. 
This is reasonable since, in colour singlet channels, $N_c$ enters only
parametrically in the relevant physical constants and observables. 
For this reason, we fix those parameters through the constraints imposed by 
the pion decay constant and the chiral (quark) condensate:
\begin{itemize}
\item{The pion decay constant $f_{\pi}$  
is evaluated in the NJL model through the following relation:
\beq
f_{\pi}^{2}=4m^2I_{\Lambda}^{(1)}\left(m\right)~~~\mathrm{where}~~~
I_{\Lambda}^{(1)}\left(m\right)
=-iN_c\int\frac{d^4p}{\left(2\pi\right)^4}\frac{\theta\left(\Lambda^2-
\vec{p}^{~2}
\right)}{\left(p^2-m^2+i\epsilon\right)^2}.
\label{fp}
\eeq
The empirical value is $f_{\pi}=92.4$ MeV.}
\item{The quark condensate becomes
\beq
\left<{\bar \psi}_u\psi_u\right>=
-4mI_{\Lambda}^{(0)}\left(m\right)
\eeq
with
\beq
I_{\Lambda}^{(0)}\left(m\right)=iN_c\int\frac{d^4p}{\left(2\pi\right)^4}\frac
{\theta\left(\Lambda^2-\vec{p}^{~2}\right)}{p^2-m^2+i\epsilon}.
\eeq
Its ``empirical'' value derived from QCD sum rules is 
\beq
\langle\bar{\psi}_u\psi_u\rangle^{1/3}\simeq\langle\bar{\psi}_d\psi_d\rangle
^{1/3}=-\left(240\pm20\right)~\mathrm{MeV}.
\eeq
}
\item{The current quark mass $m_0$ is fixed from the Gell-Mann, Oakes, Renner
(GMOR) relation:
\beq
m_{\pi}^{2}=\frac{-m_0\left<{\bar\psi}\psi\right>}{f_{\pi}^2}.
\label{gmor}
\eeq 
In the chiral limit, $m_0=0$ and $m_{\pi}=0$.}
\end{itemize}

The Goldberger-Treiman relation, which determines the pion-quark coupling
$g_{\pi}$, 
follows from the previous relations:
\beq
m=g_{\pi}f_{\pi}
\eeq
with $g_{\pi}^{2}=(4I_{\Lambda}^{(1)}(m))^{-1}$.

We will first perform all our calculations with a finite value for the 
current quark mass $m_0$, and then investigate the chiral limit, 
$m_0\rightarrow 0$.
The parameters obtained by imposing the constraints~(\ref{fp}-\ref{gmor}) are 
shown in Table~\ref{tabella}.
\begin{table}
\begin{center}
\begin{tabular}{c|c|c|c|c|c|c}
\hline
\hline
\vspace{-0.45 cm}\\
$\Lambda$ [GeV]&$G=H$[GeV$^{-2}$]&$m_0$[MeV]&$m$[MeV]&
$|\langle{\bar \psi}_u\psi_u\rangle|$$^{1/3}$[MeV]&$f_{\pi}$[MeV]&$m_{\pi}$[MeV]\\
\hline
0.78&10.3&4.5&361&259&89.6&139.3\\
\hline
\hline
\end{tabular}
\caption{Parameter set used in this work, and the corresponding physical 
quantities.}
\label{tabella}
\end{center}
\end{table}
\section{Results at finite $T$ and $\mu$}
We now extend the NJL model to finite temperature $T$ and
chemical potentials $\mu$ using the Matsubara formalism. We
consider the isospin symmetric case, with an equal number (and therefore a 
single chemical potential) of $u$ and $d$ quarks.
The quantity to be minimized at finite temperature is the thermodynamic
potential:
\bea
\Omega\left(T,\mu\right)&=&-T\sum_n\int\frac{d^3p}{\left(2\pi\right)^3}\frac 12
\mathrm{Tr}\ln\left(\frac 1T\tilde{S}^{-1}\left(i\omega_n,\vec{p}\right)\right)
\nonumber\\
&+&\frac{\sigma^2}{2G}+\frac{\left|\Delta\right|^2}{2H},
\label{omega}
\eea
where $\omega_n=(2n+1)\pi T$ are the Matsubara frequencies for fermions and 
the inverse quark propagator including the chemical potential $\mu$ is now 
defined as
\beq
\tilde{S}^{-1}\left(p^0,\vec{p}\right)=\left({{\begin{array}{ccc} \pslash-
\hat{M}-\mu\gamma_0 & \Delta\gamma_5\tau_2t_2\\-\Delta^*\gamma_5\tau_2
t_2 & \pslash-\hat{M}+\mu\gamma_0\end{array}}}\right).
\label{propmu}
\eeq
Using the identity
\beq
\mathrm{Tr}\ln\left(X\right)=\ln\det\left(X\right)
\eeq
we can evaluate the trace in (\ref{omega}) and obtain
\bea
\frac 12\mathrm{Tr}\ln\left(\frac{\tilde{S}^{-1}}{T}\left(i\omega_n,\vec{p}
\right)\right)=4\ln\left(\frac{\omega_{n}^2+{(E^+)}^2}{T^2}\right)+
4\ln\left(\frac{\omega_{n}^2+{(E^-)}^2}{T^2}\right),
\label{trace}
\eea
where we have defined $E^{\pm}=\sqrt{\left(\epsilon^{\pm}
\right)^2+\left|\Delta\right|^2}$, with  
$\epsilon^{\pm}=\epsilon\pm\mu$, $\epsilon=\sqrt{\vec{p}^{~2}+m^2}$. 
Next we evaluate the Matsubara sum in eq.~(\ref{omega}) using the 
following relation:
\beq
T\sum_{n=-\infty}^{\infty}\ln\left(\frac{\omega_{n}^2+{E^{\pm}}^2}{T^2}\right)
=E^{\pm}+2T\ln\left(1+\exp\left(-E^{\pm}/T\right)\right).
\label{sum}
\eeq
The thermodynamic potential becomes:
\bea
\Omega\left(T,\mu\right)&=&-4\int\frac{\mathrm{d}^3 p}{(2\pi^3)}\left[2T
\ln\left(1+\exp\left(-\frac{E^+}{T}\right)\right)+\right.
\label{potential}\\
&+&\left.2T\ln
\left(1+\exp\left(-\frac{E^-}{T}\right)\right)+ 
\left(E^++E^-\right)\right]\theta\left(
\Lambda^2-\vec{p}^{~2}\right)+\frac{\sigma^2}{2G}+\frac{|\Delta|^2}{2H}.
\nonumber
\eea
In eqs.~(\ref{trace})-(\ref{potential}), the effective (constituent) quark 
mass $m$ is related to the current
quark mass and the $\sigma$ field through eq.~(\ref{mass}).

The mean values for the $\sigma$ and $\Delta$ fields are determined by
minimizing the thermodynamic potential. One obtains 
the following set of coupled equations that must be
solved simultaneously in order to find the solutions for $\sigma$ and 
$|\Delta|$:
\newpage
\bea
\sigma&=&-\frac{2G}{\pi^2}\int\mathrm{d}p~p^2~\frac{m_0-\sigma}{\epsilon}\left
[\frac{\epsilon-\mu}{E^-}+\frac{\epsilon+\mu}{E^+}+\right.\\
\nonumber
&&~~~~~~~~~~~~~~~~~~~~~~~~~~~~~~-\left.2\left(\frac{\epsilon-\mu}
{\left(\exp\left(\frac{E^-}{T}\right)+1\right)E^-}+\frac{\epsilon+\mu}{\left(
\exp\left(\frac{E^+}{T}\right)+1\right)E^+}\right)\right]\\
|\Delta|&=&\frac{2H}{\pi^2}\int\mathrm{d}p~p^2~\left[\frac{
|\Delta|}{E^-}+\frac{|\Delta|}{E^+}-2\left(\frac{|\Delta|}{\left(
\exp\left(\frac{E^-}{T}\right)+1\right)E^-}+\frac{|\Delta|}{\left(
\exp\left(\frac{E^+}{T}\right)+1\right)E^+}\right)\right]
\nonumber
\label{fieldequations}
\eea
\begin{figure}
\parbox{5cm}{
\scalebox{1.2}{
\includegraphics*[49,562][321,730]{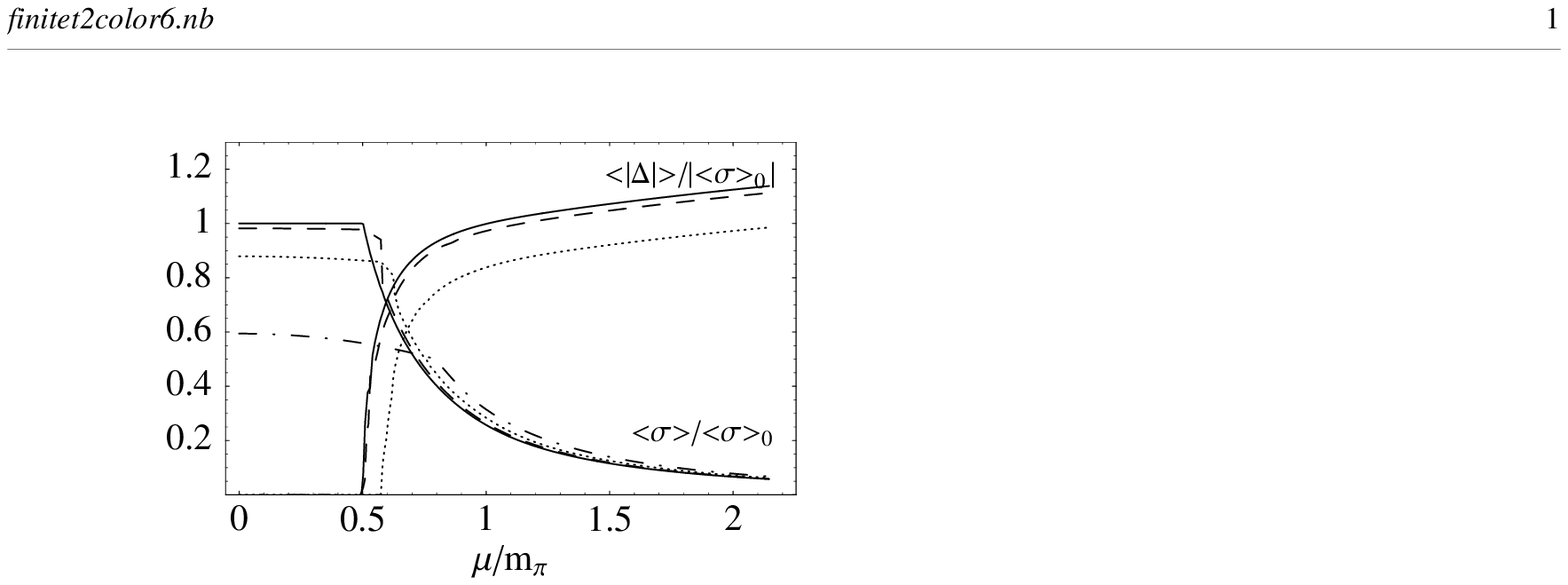}}
}\\
\parbox{15cm}{
\caption{
\footnotesize Scaled expectation values $\langle\sigma\rangle$ and $\langle|
\Delta|\rangle$ as a function of the chemical 
potential for different temperatures. Continuous lines correspond to $T=0$,
dashed lines to $T=100$ MeV, dotted lines to $T=150$ MeV and the dashed-dotted
line corresponds to $T=200$ MeV ($\langle|\Delta|\rangle=0$ in this case). 
\label{fig1}}
}
\end{figure}
In Fig.~\ref{fig1} we show our results for the scaled expectation values of the
$\sigma$ and $\Delta$ fields as a function of the chemical potential for 
different temperatures. One observes that at $T=0$ the system undergoes a 
second 
order phase transition at a critical chemical potential $\mu_c=m_{\pi}/2$, as 
predicted by general arguments. The value of the pion mass that we consider 
here is the one evaluated in the model and shown in 
Table~\ref{tabella}. So this model exhibits diquark condensation at chemical 
potentials larger than $\mu_c$, where the value of the chiral 
condensate is correspondingly reduced. At $T=0$, $\Delta$ is always non-vanishing for $\mu>\mu_c$: the diquark phase persists for large $\mu$. For temperatures
$T\gtrsim 200$ MeV, on the other hand, the diquark condensate vanishes even
for large chemical potentials.

The chiral effective Lagrangian approach~\cite{Kogut:2000ek} predicts the following behaviour
for the diquark condensate as a function of the chemical potential at 
$\mu>\mu_c$:
\beq
\frac{\langle\psi\psi\rangle}{|\langle\bar{\psi}\psi\rangle_0|}=
\frac{\langle|\Delta|\rangle}{|\langle\sigma\rangle_0|}=\sqrt{1-\left(\frac
{m_{\pi}}{2\mu}\right)^4},
\label{chirald}
\eeq
which means that $\langle|\Delta|\rangle$ should reach the vacuum expectation 
value of the (scaled) chiral
condensate asymptotically as $\mu\rightarrow\infty$. In the NJL model, the 
scale of variation for $\mu>\mu_c$ is set by the momentum cutoff 
$\Lambda$. As a consequence, $|\Delta(\mu)|$ increases until 
$\mu\sim\Lambda$ (corresponding to $\mu/m_{\pi}\sim 5)$. For larger values of 
$\mu$ the relevant interactions become
weaker and $|\Delta|$ tends to decrease with $\mu$. This feature is an 
artifact, however, since the 
applicability of the NJL model is limited to energy and momentum scales 
below $\Lambda$. For chemical potentials 
smaller than the cutoff scale the agreement between NJL and chiral Lagrangian 
calculations is excellent, as expected. At very large chemical potential, 
perturbative gluon exchange presumably takes over, with decreasing interaction
strength as $\mu$ increases.

\begin{figure}
\begin{center}
\parbox{5cm}{
\scalebox{0.5}{
$\!\!\!\!\!\!\!\!\!\!\!\!\!\!\!\!\!\!\!\!\!\!\!\!\!\!\!\!\!\!\!\!\!\!\!\!\!\!\!\!\!\!\!\!\!\!\!\!$\includegraphics*[88,219][658,630]{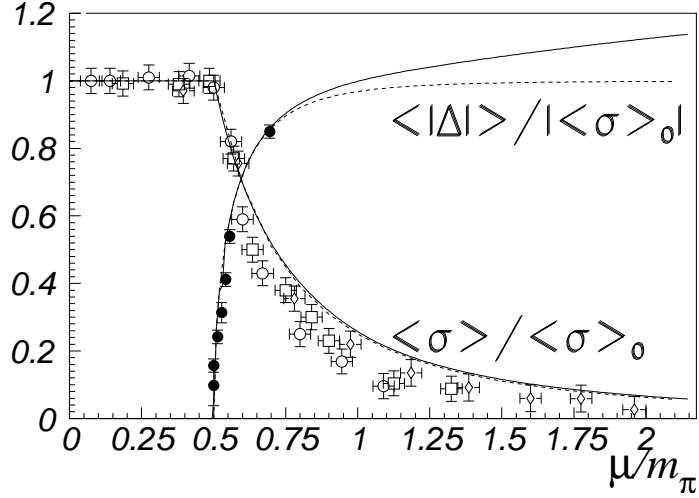}}}
\end{center}
\parbox{15cm}{
\caption{
\footnotesize Scaled $\langle\sigma\rangle$ and $\langle|\Delta|\rangle$ as a 
function of the chemical potential at $T=0$: our results (solid lines) are 
compared to the lattice data taken from ref.~\cite{Hands:2001ee}. The different
symbols (open circles, squares and diamonds) for the chiral condensate correspond to different values for the quark 
masses. The dashed lines are the predictions from chiral effective field 
theory~\cite{Kogut:2000ek}.
\label{latticechiral}}
}
\end{figure}

In Fig.~\ref{latticechiral} we show a comparison of our results for the scaled
chiral and diquark condensates at $T=0$ as a function of the chemical 
potential, with lattice data taken from ref.~\cite{Hands:2001ee}. These data 
have been obtained by studying two-colour QCD with staggered fermions in the
adjoint representation. It was found that the positive determinant sector 
behaves
like a two-flavour theory. As we can see, the agreement of our results with 
lattice data is remarkable. The dashed lines are the predictions from chiral
effective field theory.

\begin{figure}
\hspace{-.05\textwidth}
\begin{minipage}[t]{.48\textwidth}
\parbox{5cm}{
\scalebox{0.95}{
\includegraphics*[99,572][327,727]{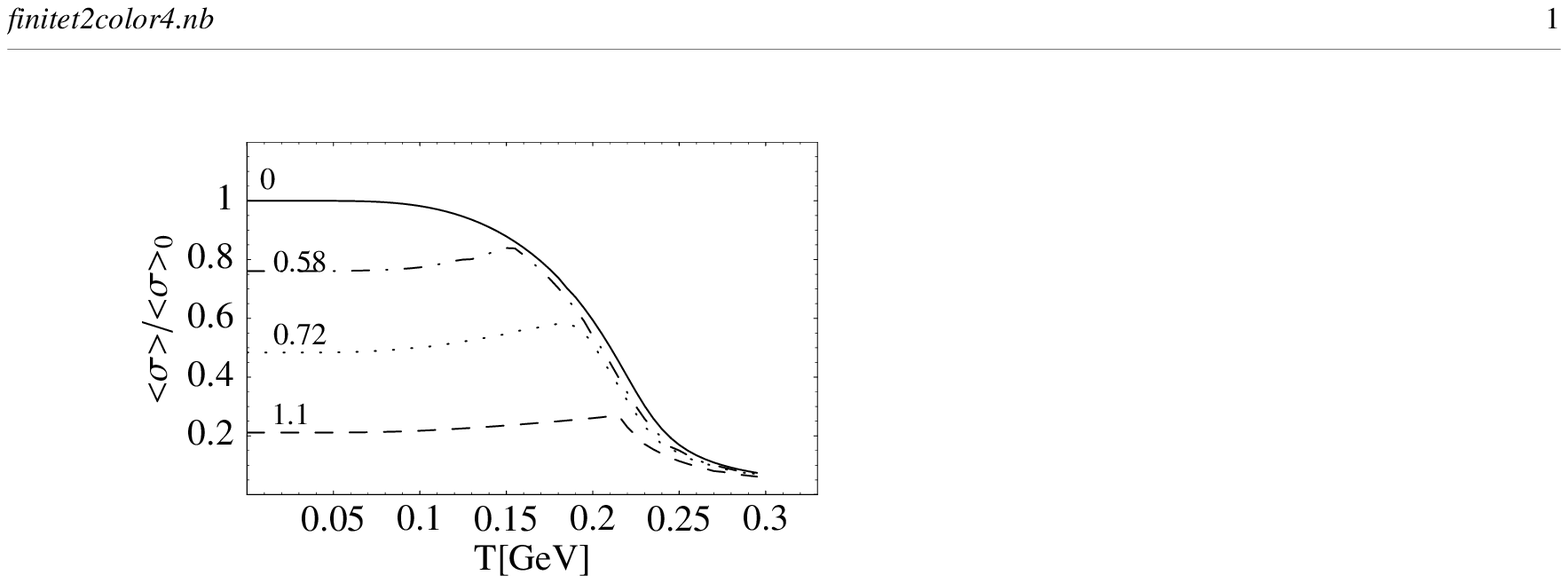}\\}}
\centerline{(a)}
\end{minipage}
\hspace{.02\textwidth}
\begin{minipage}[t]{.48\textwidth}
\parbox{5cm}{
\scalebox{0.95}{
\includegraphics*[89,572][327,727]{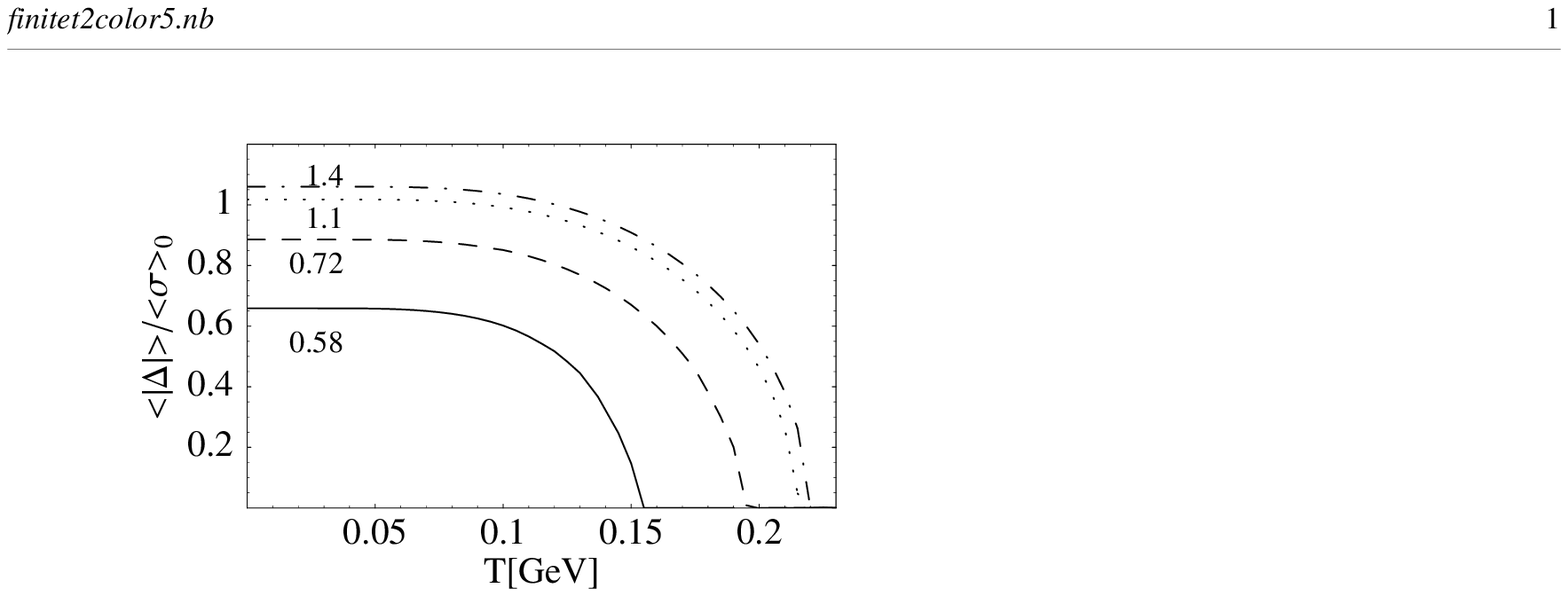}\\}}
\centerline{(b)}
\end{minipage}
\parbox{15cm}{
\caption{
\footnotesize Scaled $\langle\sigma\rangle$ (a) and $\langle|\Delta|\rangle$ (b) as a function of
temperature for different values of $\mu/m_{\pi}$.}
\label{ab}}
\end{figure}

\begin{figure}
\begin{center}
\mbox{\epsfig{file=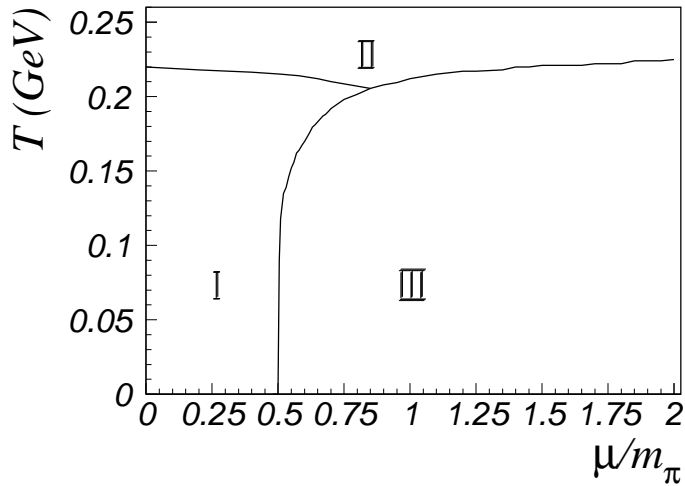, height=.35\textheight}}
\end{center}
\vskip -0.8cm
\parbox{15cm}{
\caption{
\footnotesize Phase diagram in the NJL model with two colours. The zone I 
is a region in which chiral symmetry is spontaneously broken, and  
$\langle|\Delta|\rangle=0$; in region
II chiral symmetry is restored, and again $\langle|\Delta|\rangle=0$; region III
is the superfluid phase in which $\langle|\Delta|\rangle\neq 0$.  
\label{phase}}
}
\end{figure}

In Fig.~\ref{ab} we show the scaled $\langle\sigma\rangle$ and $\langle|
\Delta|\rangle$ as a function
of the temperature for different values of the chemical potential. In this way
we find, as a function of the chemical potential, the critical 
temperature of the phase transition, so that we can draw the phase diagram of 
two-colour QCD as modelled in the NJL model. We show it in Fig.~\ref{phase}.  
At very small chemical potentials we have a transition
from a system in which chiral symmetry is spontaneously broken to a system 
where it is restored (from region I to region II) with $\langle|\Delta|\rangle=
0$ in both phases. Region III is the superfluid phase with 
$\langle|\Delta|\rangle\neq 0$.

\begin{figure}
\parbox{5cm}{
\scalebox{.9}{
\includegraphics*[49,492][421,730]{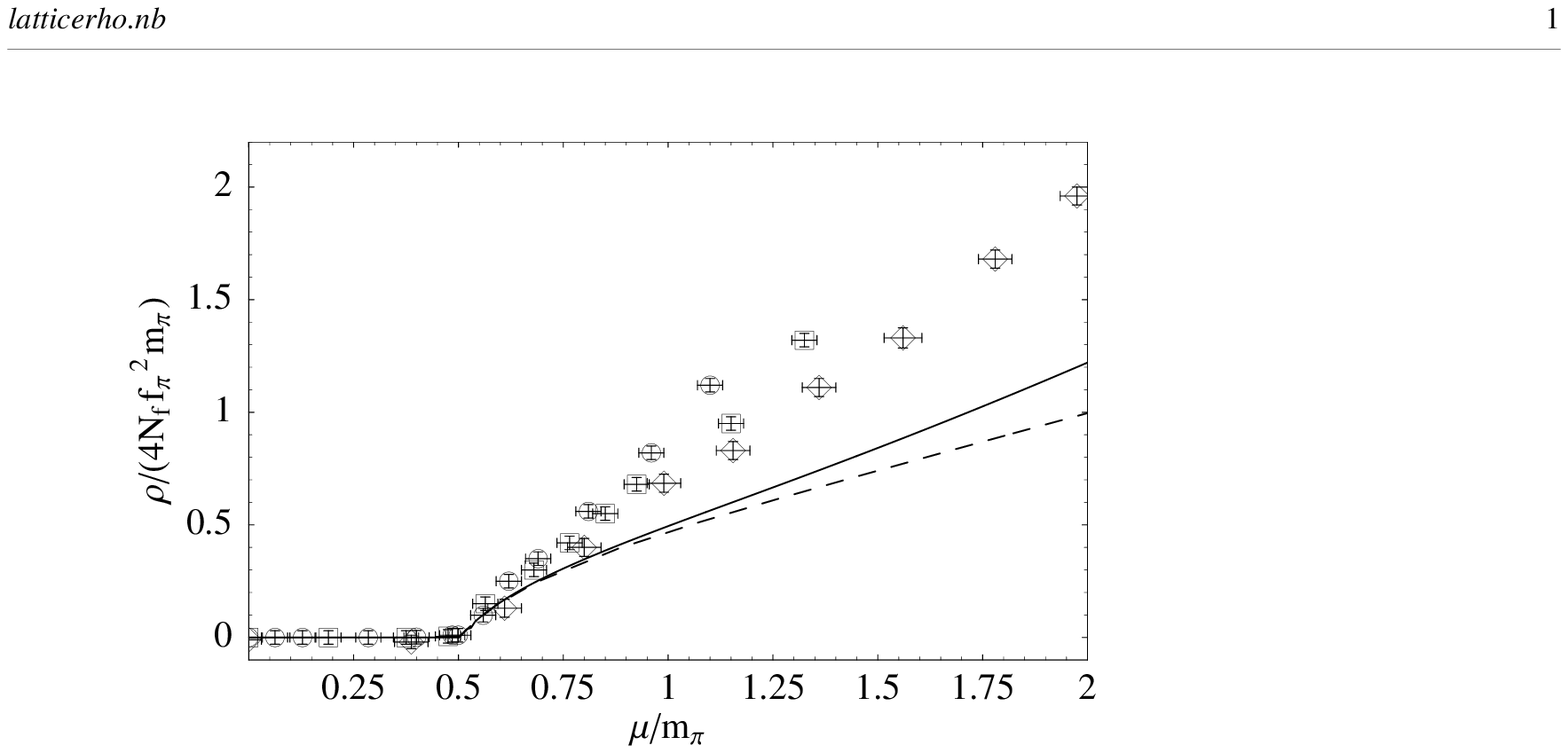}}
}\\
\parbox{15cm}{
\caption{
\footnotesize Scaled baryonic density as a function of the chemical potential
at $T=0$ (Continuous line). The lattice data are taken from 
ref.~\cite{Hands:2001ee}. 
The different symbols correspond to different values for the quark masses.
The dashed line is the prediction from chiral effective field 
theory~\cite{Kogut:2000ek}.
\label{latticerho}}
}
\end{figure}

An interesting quantity is the baryonic density
\beq
\rho=-\frac{\partial\Omega(T,\mu)}{\partial\mu}.
\eeq
The lattice data of ref.~\cite{Hands:2001ee} show a scaled baryonic density
defined as:
\beq
\tilde{\rho}=\frac{\rho}{4N_ff_{\pi}^{2}m_{\pi}}.
\label{rhotilde}
\eeq
Leading order chiral effective field theory~\cite{Kogut:2000ek} gives the 
following behaviour at $\mu>\mu_c$:
\beq
\tilde{\rho}=\frac{\mu}{2m_{\pi}}\left(1-\left(\frac{m_{\pi}}{2\mu}\right)^4
\right).
\eeq
Fig.~\ref{latticerho} presents our results for the scaled baryonic density
(\ref{rhotilde}) as a function of the chemical potential at zero temperature, 
in comparison with the lattice data for the same quantity.
Our results are in good agreement with lattice data at moderate
chemical potentials, while for large chemical potentials the baryon density is
underestimated. This difference may be caused by the mean-field approximation.
Correlations between quasiparticles, not covered by this approximation, tend to
become increasingly important with growing density. 

\subsection{Pion and scalar diquark properties}
This Section presents our results for the masses of the (pseudo)-Goldstone 
bosons, namely the pion, the scalar diquark and the corresponding antidiquark.

In order to evaluate the masses of the bosonic fields, we expand the 
effective action 
\beq
\mathcal{S}_{eff}=-\int\mathrm{d}^4x\left[\frac{\sigma^2+\vec{\pi}^2}
{2G}+\frac{\Delta\Delta^*}{2H}\right]-\frac {i}{2}tr\int\mathrm{d}^4x\ln
\left(S^{-1}\left(x\right)\right).
\eeq
in a power series of the meson and diquark fields around their mean field
values. The second-order term of this expansion identifies the mass
spectrum of mesons and diquarks.
The resulting effective action in momentum space has the following form:
\beq
\mathcal{S}^{(2)}_{eff}\left(\sigma,\vec{\pi},\Delta,\Delta^*\right)=
-\frac{\sigma^2+\vec{\pi}^2}{2G}-\frac{\Delta\Delta^*}{2H}+\frac{i}{4}tr
\int\frac{\mathrm{d}^4p}{\left(2\pi\right)^4}\left[\tilde{S}_0A\tilde{S}_0A\right]
\label{S2}
\eeq
where $\tilde{S}_0$ is the Nambu-Gorkov propagator~(\ref{propmu}) 
evaluated at the mean field values for the bosonic fields,
and $A$ is a matrix defined in the following way:
\beq
A=\left({{\begin{array}{ccc}
\sigma+i\gamma_5\vec{\pi}\cdot\vec{\tau} & \Delta\gamma_5\tau_2t_2\\
-\Delta^*\gamma_5\tau_2t_2&\sigma-i\gamma_5\vec{\pi}\cdot\vec{\tau}\end{array}}
}\right)
\label{A}
\eeq
(see also~\cite{Blaschke:2004cs}).
By analyzing the second order action~(\ref{S2}), one observes that 
mixing terms arise, at $\mu>\mu_c$, between the $\sigma$, $\Delta$ and
$\Delta^*$ fields: these terms are proportional to $|\Delta|$, 
and the mixing occurs because the presence of a nonzero diquark condensate 
spontaneously breaks the baryon number symmetry. This feature was already
found in~\cite{Kogut:2003ju}. The mass matrix turns out to
have the following form:
\bea
M=\left({{\begin{array}{cccc}
\frac{{\partial}^2\mathcal{S}^{(2)}_{eff}}{\partial\vec{\pi}^2}&0&0&0
\vspace{.2cm}\\
0&\frac{{\partial}^2\mathcal{S}^{(2)}_{eff}}{\partial{\sigma}^2}&\frac{{\partial}^2\mathcal{S}^{(2)}_{eff}}{\partial\sigma
\partial\Delta}&\frac{{\partial}^2\mathcal{S}^{(2)}_{eff}}{\partial\sigma
\partial{\Delta}^*}
\vspace{.2cm}\\
0&\frac{{\partial}^2\mathcal{S}^{(2)}_{eff}}{\partial\Delta\partial\sigma}&
\frac{{\partial}^2\mathcal{S}^{(2)}_{eff}}{\partial{\Delta}^2}&\frac{{\partial}^2\mathcal{S}^{(2)}_{eff}}{\partial\Delta
\partial{\Delta}^*}
\vspace{.2cm}\\
0&\frac{{\partial}^2\mathcal{S}^{(2)}_{eff}}{\partial{\Delta}^*\partial\sigma}&\frac{{\partial}^2\mathcal{S}^{(2)}_{eff}}
{\partial{\Delta}^*\partial\Delta}&\frac{{\partial}^2\mathcal{S}^{(2)}_{eff}}
{\partial{{\Delta}^*}^{2}}
\end{array}}}\right),
\label{Mmatrix}
\eea
and the masses of the various modes are found by solving the equation
\beq
\det\left(M\right)=0.
\label{detM}
\eeq
Evidently the pion fields do not mix with the others, while the $\sigma$,
the diquark and the antidiquark fields mix in the phase with $|\Delta|\neq0$.

\begin{figure}
\begin{center}
\mbox{\epsfig{file=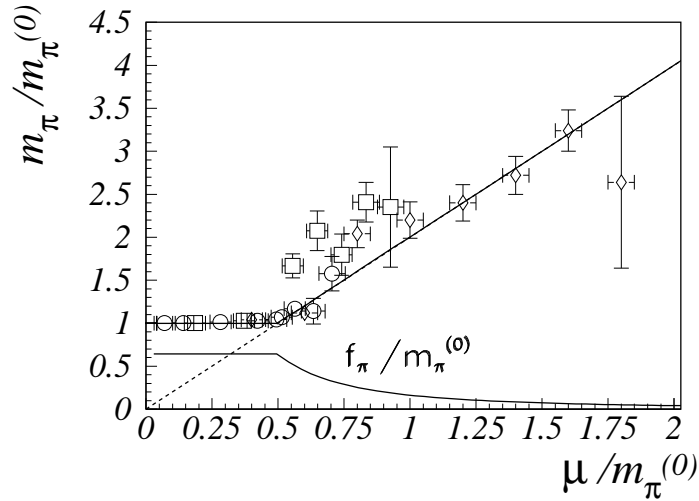, height=.35\textheight}}
\end{center}
\vskip -0.8cm
\parbox{15cm}{
\caption{
\footnotesize Scaled pion mass as a function of $\mu/m_{\pi}^{(0)}$
at $T=0$ (continuous line). The lattice data are taken from 
ref.~\cite{Hands:2001ee} and have been rescaled in order to show dimensionless
quantities.  
The different symbols correspond to different values for the quark masses. 
The dashed line is $m_{\pi}=2\mu$, as predicted in leading-order chiral
effective field theory~\cite{Kogut:2000ek}. Also shown is the (scaled) pion 
decay constant $f_{\pi}/m_{\pi}^{(0)}$ and its evolution with increasing $\mu$.
\label{latticempai}}
}
\end{figure}

The behaviour of the scaled pion mass as a function of 
the chemical potential is shown in Fig.~\ref{latticempai}, in comparison to
the lattice data.
The pion mass increases linearily with the chemical potential at $\mu>\mu_c$.
This behaviour was anticipated in the calculations by Kogut
{\it et al.}~\cite{Kogut:2000ek}. They in fact predicted
for $m_{\pi}$ the following behaviour at $\mu>\mu_c$:
\beq
m_{\pi}=2\mu,
\label{mpaichir}
\eeq
as indicated by the dashed line in Fig.~\ref{latticempai}.
Our result is in very good
agreement with both the lattice data and the predictions using the 
leading-order chiral effective Lagrangian.

\begin{figure}
\parbox{5cm}{
\scalebox{1.2}{
\includegraphics*[45,572][320,727]{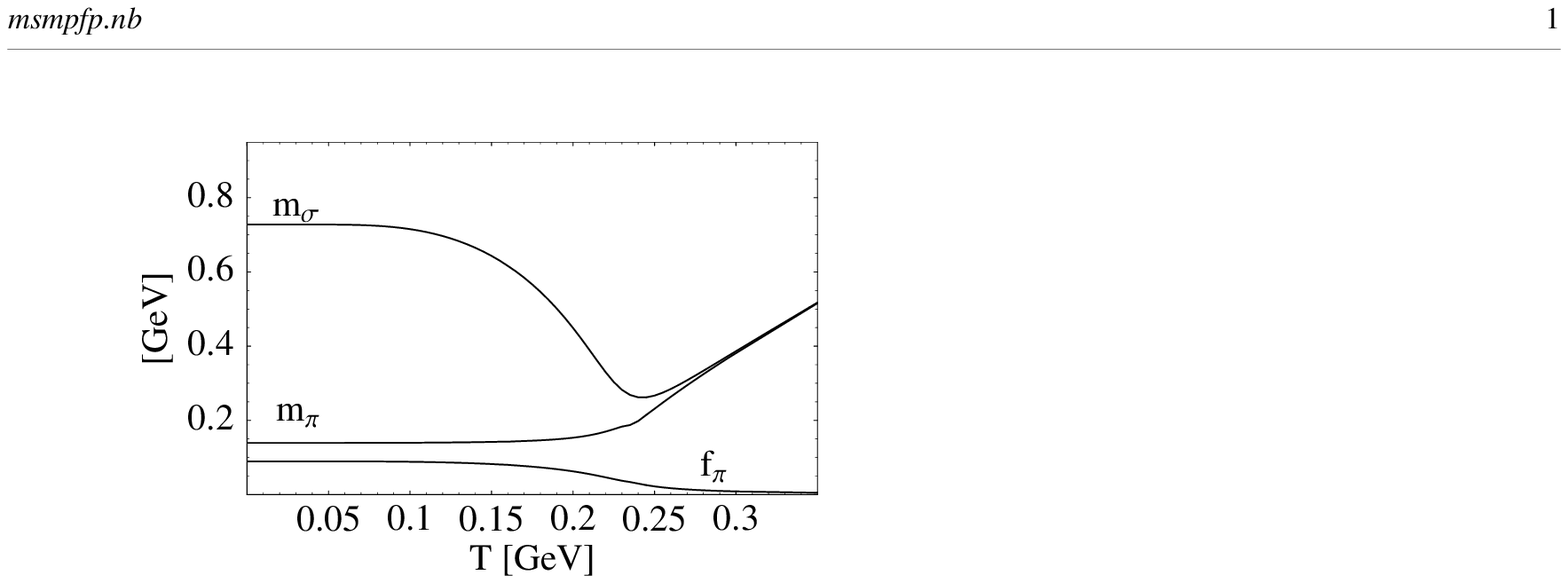}}}\\
\parbox{15cm}{
\caption{
\footnotesize Pion mass, $\sigma$ boson mass and pion decay constant as a 
function of temperature at $\mu=0$.}
\label{fig8}}
\end{figure}

The behaviour of the pion and $\sigma$ masses and of the pion decay constant 
as functions of temperature at $\mu=0$ is shown in Fig.~\ref{fig8}. At
temperatures $T$ exceeding the critical $T_c$ for the chiral transition
at which $\langle\sigma\rangle$ tends to zero, $m_{\sigma}$ becomes equal to 
the pion mass and both masses rise continuously with increasing $T$. The
pion decay constant tends to zero at the same time. 

\begin{figure}
\parbox{5cm}{
\scalebox{1.2}{
\includegraphics*[59,562][421,725]{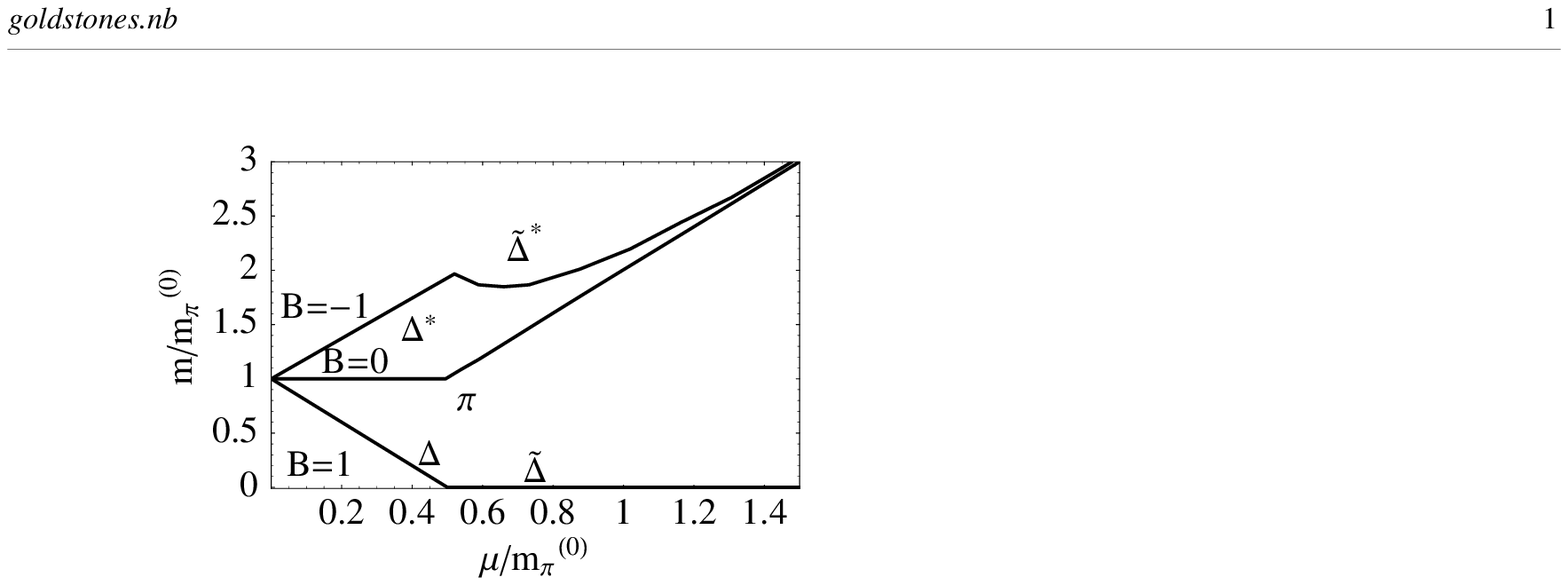}}}\\
\parbox{15cm}{
\caption{\footnotesize Spectrum of pions and diquarks/antidiquarks as a 
function of the (scaled) chemical potential at zero temperature.
\label{mdiq}}}
\end{figure}

Next, consider the other two bosonic modes of the theory: the
scalar diquark and its antidiquark. The behaviour of their masses at finite 
chemical potential is shown in Fig.~\ref{mdiq} in comparison to the pion 
mass: at $\mu=0$ they are all degenerate, as predicted on the basis of general
arguments,  but they behave in different ways as 
the chemical potential increases: for $\mu<\mu_c=m_{\pi}^{(0)}/2$ the pion, 
which 
does not carry baryon charge, is not affected by $\mu$, while the diquark 
and antidiquark masses are shifted according to their baryon number 
$B=\pm 1$. 
They follow in fact the behaviour observed also in chiral effective field 
theory~\cite{Kogut:2000ek}: 
\beq
m_{\Delta}=m_{\pi}-2\mu,~~~~~~~~~~~~~~~~~~m_{\Delta^*}=m_{\pi}+2\mu.
\eeq
For $\mu>\mu_c$, the appearance of the diquark condensate spontaneously breaks 
the baryon number symmetry. The scalar modes (diquark, antidiquark and sigma) 
get mixed. The new eigenmodes are linear combinations of the original
quasiparticle states. By solving 
eq.~(\ref{detM}) we find the masses of the new orthogonal modes. 
One of them, which we denote by 
$\tilde{\Delta}$, is massless and can be
identified with the true Goldstone boson of the theory, corresponding to the
spontaneous breaking of the baryon number ($U(1)$) symmetry. 
The other two modes are
massive. One of them, which we denote by $\tilde{\Delta}^*$, follows
the behaviour derived in the paper by Kogut {\it et al.}:
\beq
M_{\tilde{\Delta}^*}=2\mu\sqrt{1+3\left(m_{\pi}/2\mu\right)^4}.
\eeq
\section{Chiral limit~\label{chiral}}
In the chiral limit $m_0\rightarrow 0~(m_{\pi}\rightarrow 0)$, and at $\mu=0$, 
the thermodynamic 
potential~(\ref{potential}) (with $G=H$) is a function only of 
$\sigma^2+|\Delta|^2$, as 
already mentioned. This is a natural outcome once the relation between the 
coefficients $G$ and $H$ is fixed through the Fierz transformation of the 
colour current-current interaction
(see eq.~(\ref{coeff})). As a result, $\Omega$ is invariant under the rotation
which connects the chiral and the diquark condensate along the circle 
$\sigma^2+|\Delta|^2$=const. Because of this 
symmetry, the chiral condensate is indistinguishable from the diquark 
condensate for $m_0=\mu=0$, so that a state with finite $\langle\sigma\rangle$ 
can always be transformed into a state with finite 
$\langle|\Delta|\rangle$ and $\langle\sigma\rangle=0$. The phases with 
spontaneously broken chiral and baryon number symmetries are degenerate in this
limit.

\begin{figure}
\hspace{-.05\textwidth}
\begin{minipage}[t]{.48\textwidth}
\parbox{5cm}{
\scalebox{0.95}{
\includegraphics*[89,472][327,727]{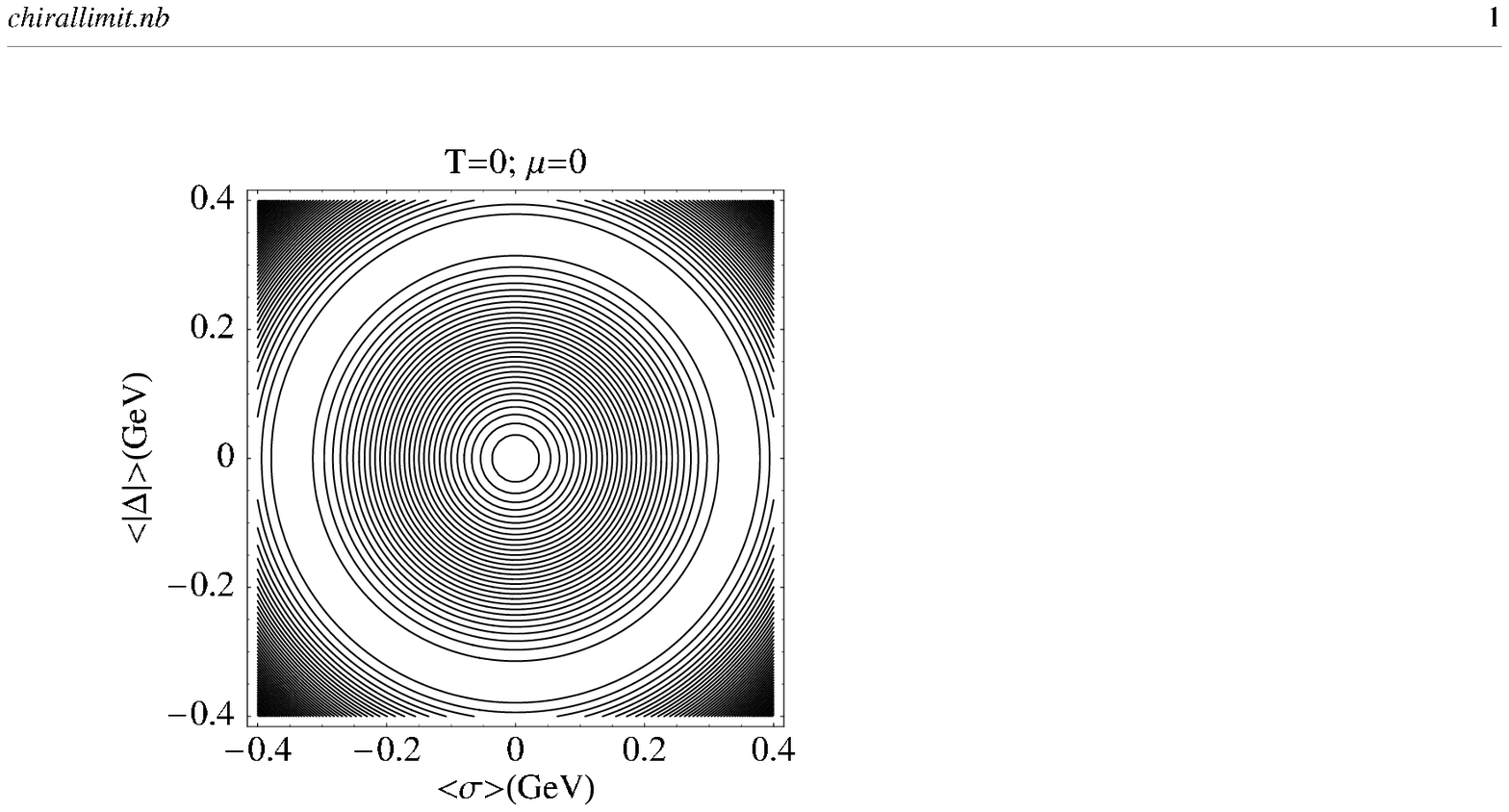}\\}}
\centerline{(a)}
\end{minipage}
\hspace{.02\textwidth}
\begin{minipage}[t]{.48\textwidth}
\parbox{5cm}{
\scalebox{0.95}{
\includegraphics*[89,472][327,727]{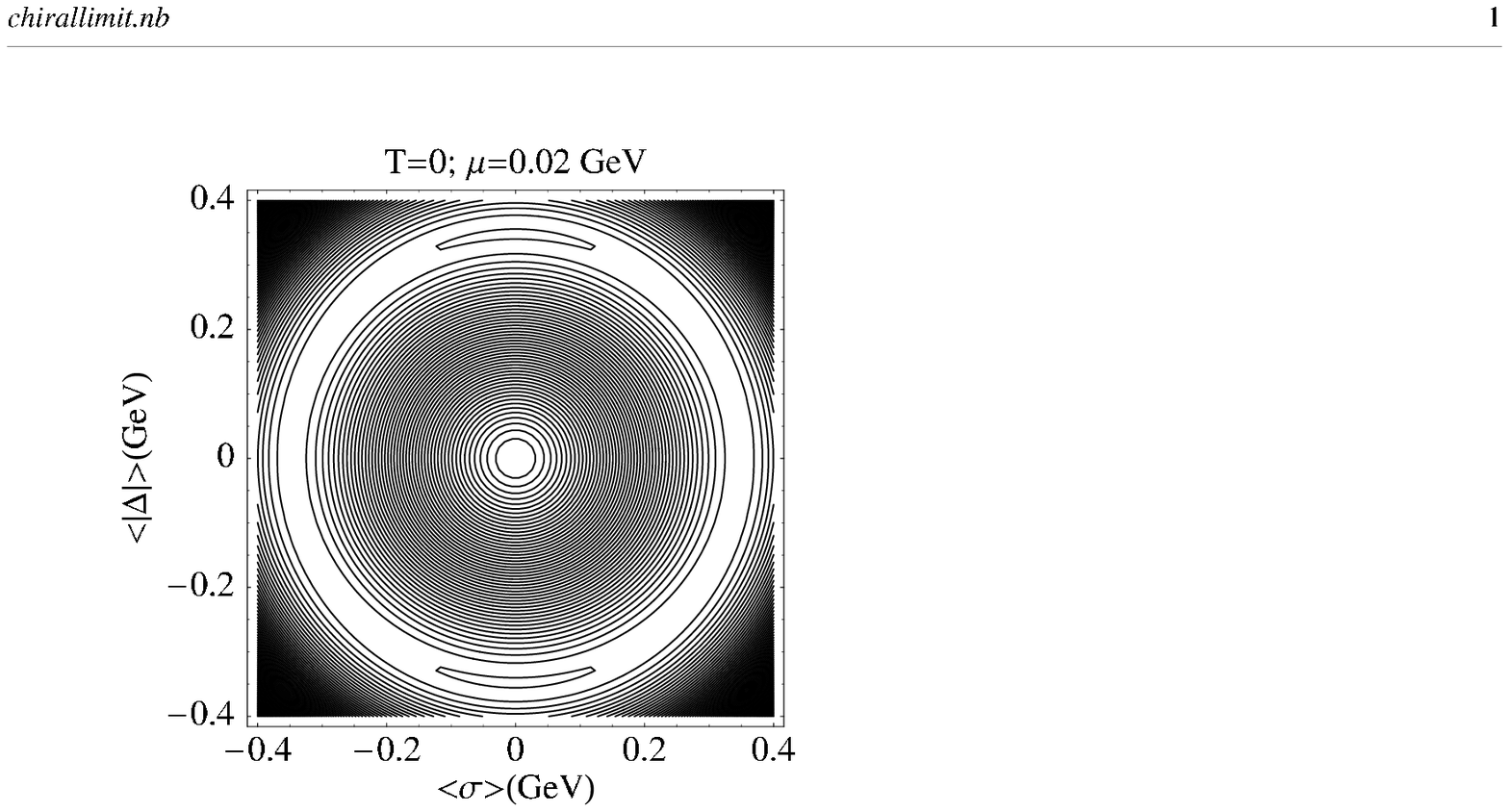}\\}}
\centerline{(b)}
\end{minipage}
\parbox{15cm}{
\caption{
\footnotesize Contour plots of the thermodynamic potential 
in the chiral limit ($m_0=0$) as a function of 
$\sigma$ and $\Delta$ for $T=\mu=0$ (a) and $T=0$ and $\mu=20$ MeV (b).}
\label{contour}}
\end{figure}

As soon as the chemical potential takes a finite value, the favourable phase 
is the one with a non-zero diquark condensate and zero chiral condensate. 
This
is evident from Fig.~\ref{contour} which shows the contour plots of the 
thermodynamic potential as a function of $\sigma$ and $|\Delta|$. In the
left panel we have $T=\mu=0$ and the rotational invariance is evident. In the
right panel we have introduced a very small chemical potential, which is 
nevertheless
sufficient to break the rotational invariance along $R^2=\langle\sigma\rangle^2
+\langle|\Delta|\rangle^2$ and favour the phase in which
$\langle\sigma\rangle=0$ and $\langle|\Delta|\rangle\neq 0$.

\begin{figure}
\begin{center}
\mbox{\epsfig{file=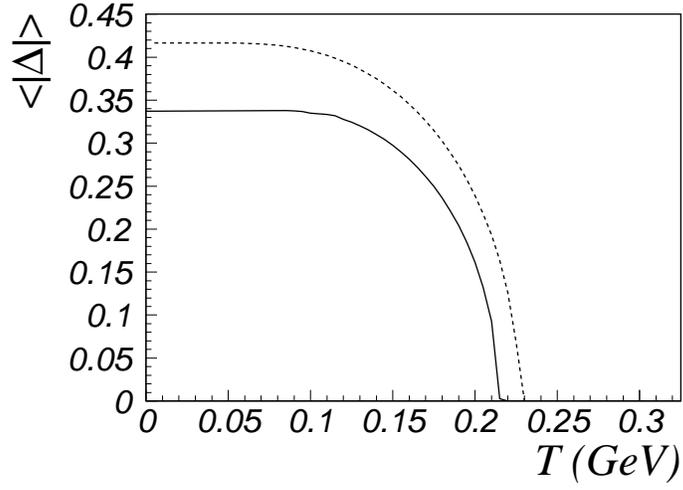, height=.35\textheight}}
\end{center}
\vskip -0.8cm
\parbox{15cm}{
\caption{
\footnotesize Mean field value of the $|\Delta|$ field as a function 
of temperature for $\mu=0$ (continuous) and $\mu=350$ MeV 
(dashed).}
\label{deltamu}}
\end{figure}

Minimizing the thermodynamic potential of the system, one finds the mean-field
values of the chiral and diquark condensates. Our results in Fig.~\ref{deltamu}
display $\langle|\Delta|\rangle$ as a function of temperature for different
chemical potentials. The chiral condensate is always equal to zero in those 
cases.

\begin{figure}
\begin{center}
\mbox{\epsfig{file=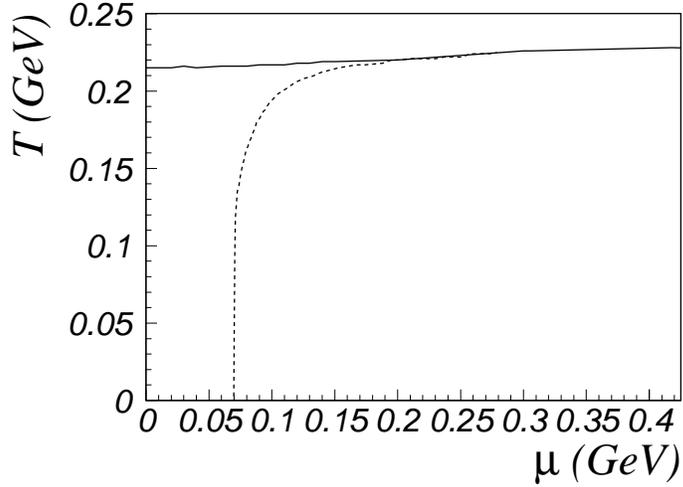, height=.35\textheight}}
\end{center}
\vskip -0.8cm
\parbox{15cm}{
\caption{
\footnotesize Comparison between the phase diagram of two-colour QCD  
in the chiral limit (continuous) and for bare quark mass $m_0\neq 0$ (dashed).
\label{chiralphase}}
}
\end{figure}

In Fig.~\ref{chiralphase} the phase diagram of the two-colour NJL model in the
chiral limit is compared to the one using a finite value of the bare quark mass
$m_0$. As one can see, the phase boundaries for $m_0=0$ and $m_0\neq 0$ become
identical at large chemical potentials, whereas at small $\mu$ they show a 
qualitatively different behaviour. In the exact chiral limit there are only two
phases in the theory: the superfluid phase with $\langle|\Delta|\rangle\neq 0$
and the high-temperature phase with $\langle|\Delta|\rangle=0$, separated by
a critical temperature of about 0.2 GeV.

Consider next the pion and diquark masses in the chiral limit and their 
variations with increasing chemical potential. The chiral condensate is always
equal to zero in this limit. Consequently, 
the $\tilde{\Delta}$ mode is a true Goldstone boson and its mass is always
equal to zero, while
the $\tilde{\Delta}^*$ and pion 
masses are 
degenerate. Explicit symmetry breaking by a finite chemical potential lets
these masses scale as $m_{\tilde{\Delta}^*}=m_{\pi}=2\mu$. The degeneracy of 
$\tilde{\Delta}^*$ and
$\pi$ is removed as soon as a small non-zero quark mass $m_0$ is introduced.
This also gives a finite mass to the $\tilde{\Delta}$ mode, which is again
equal to zero above $\mu_c=m_{\pi^{(0)}}/2$.

\begin{figure}
\parbox{6cm}{
\scalebox{1.15}{
\includegraphics*[29,572][327,727]{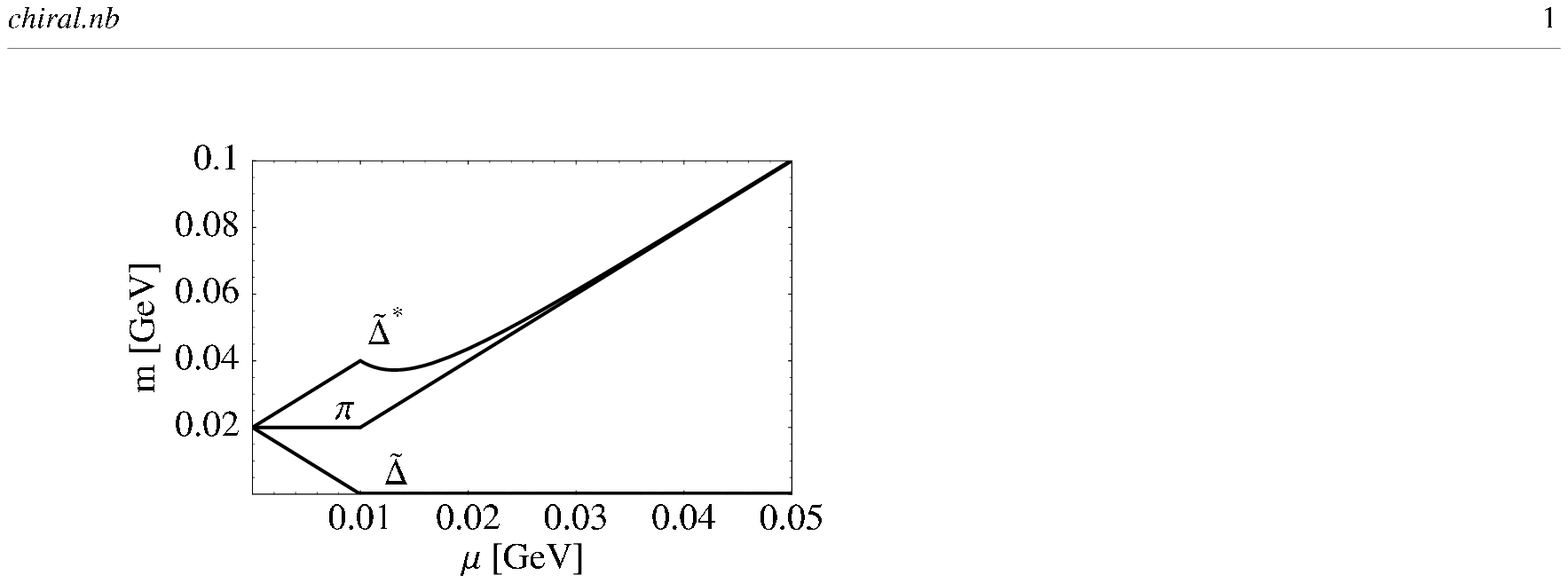}}}\\
\parbox{15cm}{
\caption{
\footnotesize Pion and diquark/antidiquark masses approaching the chiral limit 
($m_0\simeq 0.1$ MeV). In the exact chiral limit,
$m_{\tilde{\Delta}^*}=m_{\pi}=2\mu$ and $m_{\tilde{\Delta}}\equiv 0$.
\label{chiralmasses}}}
\end{figure}

Fig.~\ref{chiralmasses} illustrates this situation for a very small value 
of $m_0$ ($\sim 0.1$ MeV). The critical value $\mu_c$ of the chemical potential
is identified as $\mu_c=m_{\pi}^{(0)}/2$, as discussed previously, but now of 
course with a very small value of the vacuum pion mass $m_{\pi}^{(0)}$. As the
limit $m_0\rightarrow 0$ is approached, $m_{\pi}^{(0)}\rightarrow 0$ and 
$\mu_c\rightarrow 0$: the low-temperature system is always in the superfluid 
phase for any value of $\mu$. At $\mu=0$ we recover the exact Pauli-G\"ursey
symmetry, with vanishing pion and diquark masses.

\section{Conclusions}
We have investigated a two-colour and two-flavour Nambu and Jona-Lasinio model
at finite temperature and finite baryon chemical potential, with the primary
aim of exploring the capability of such a model to reproduce the thermodynamics
of $N_c=2$ lattice QCD. The starting point is the assumption that gluon 
dynamics can be integrated out and reduced to a local interaction between quark
colour currents. By Fierz rearrangement, this implies a one-to-one 
correspondence between interactions in colour singlet quark-antiquark and 
diquark channels (the Pauli-G\"ursey symmetry).

The resulting spontaneous (dynamical) symmetry breaking pattern identifies
pseudoscalar Goldstone bosons (pions) and scalar diquarks as the 
thermodynamically active quasiparticles. The successful comparison with 
$N_c=2$ lattice data indicates that this simple NJL model does indeed draw a 
remarkably realistic picture of the quasiparticle dynamics emerging from 
$N_c=2$ QCD, even though the original local colour gauge symmetry of QCD has
been reduced to a global colour $SU(2)$ symmetry in the NJL quasi-particle 
model. We note that colour (triplet) quark-antiquark modes which are the
remnants of gluon degrees of freedom in this model, are far removed from the
low-energy spectrum. Poles of the respective Bethe-Salpeter amplitudes appear 
at mass scales several times the NJL cutoff scale~\cite{mt}.

We confirm that a diquark condensate develops at chemical potentials 
$\mu>\mu_c=m_{\pi}/2$. The correlated evolution of the chiral and diquark
condensates with increasing $\mu$, as observed in $N_c=2$ lattice QCD, is very
well reproduced. Had we started from NJL four-point interactions with 
independent, arbitrary coupling strengths in quark-antiquark and diquark 
channels, the condensate pattern would have been quite different.
It appears that modelling the low-energy dynamics of $N_c=2$ QCD is already
done surprisingly well when using just a colour current-current interaction 
with a single strength parameter.

The calculated baryon density, obtained by taking the derivative of the
thermodynamic potential with respect to the chemical potential, describes
the corresponding lattice results well in the range $\mu<2\mu_c$. Deviations
occur at larger $\mu$ which presumably indicate the increasing importance
of correlations between quasiparticles beyond the mean-field approximation.

The NJL model also permits an instructive study of the way in which this system
behaves in the chiral limit which is not directly accessible in lattice
computations. In particular, the limits of vanishing quark mass and vanishing 
baryon chemical potential do not commute, as expected, and have to be handled 
with care.

The low-energy physics of QCD differs qualitatively between $N_c=2$ and $N_c=3$
because of the very different nature of the baryonic quasiparticles in these
two theories. Nevertheless, the success of the present studies encourages
further extended investigations also for $N_c=3$ thermodynamics, using NJL
type quasiparticle approaches above the critical temperature for deconfinement,
in close contact with lattice QCD simulations.

\bigskip
\bigskip
\bigskip
\noindent
We thank Pietro Faccioli, Jiri Hosek, Georges Ripka and Michael Thaler for
stimulating discussions and valuable comments.
\section*{Appendix}
We start from the colour current interaction (\ref{lc}) and 
show that performing a Fierz transformation we obtain the Lagrangian
(\ref{lag}) with the coupling coefficients related by (\ref{coeff}).

In order to demonstrate this identity for the coefficient of the scalar diquark
interaction ($H$) we must Fierz transform this interaction into the $qq$ 
channel, while for $G$ we must Fierz transform into the $\bar{q}q$ channel.
 
Let us start with $H$; we rewrite the interaction term of eq.~(\ref{lc}) and
keep track explicitly of all colour, flavour and Dirac indices:

\bea
\mathcal{L}_{int}^c&=&-G_c\sum_{a=1}^3\left(\bar{\psi}\gamma_{\alpha}t^a
\psi
\right)^2
\label{explicit}
\\
\nonumber
&=&-G_c\sum_{a=1}^3\left[\bar{\psi}_{i,p,\mu}\psi_{j,q,\nu}\bar{\psi}
_{k,r,\rho}\psi_{l,s,\sigma}\left(\gamma_{\alpha}\right)_{\mu\nu}
\left(\gamma^{\alpha}\right)_{\rho\sigma}\left(t_a\right)_{ij}\left
(t_a\right)_{kl}\delta_{pq}\delta_{rs}\right]
\eea
with:
\bea
\nonumber
i,j,k,l~~~~~~~~~~~~~~&\rightarrow&~~~~~~~~~~~~~~\mathrm{colour~indices}\\
\nonumber
p,q,r,s~~~~~~~~~~~~~~&\rightarrow&~~~~~~~~~~~~~~\mathrm{flavour~indices}\\
\nonumber
\mu,\nu,\rho,\sigma~~~~~~~~~~~~~~&\rightarrow&~~~~~~~~~~~~~~
\mathrm{Dirac~indices}.
\nonumber
\eea
We start by performing the Fierz transformation for the flavour indices 
using the following relation
\beq
\delta_{pq}\delta_{rs}=\frac 12\sum_{b=0}^3\left(\tau_b\right)_{pr}
\left(\tau_b\right)_{sq},
\eeq
where we have defined 
\beq
\tau_0=\left({{
\begin{array}{cc}
1&0\\
0&1\end{array}}}\right)~~~~~~\mathrm{and}~~~~~~\tau_b~=~\mathrm{Pauli~ 
matrices~with}~~~b=1,2,3~~~,
\eeq
thus obtaining
\bea
\mathcal{L}_{int}^c=
=-\frac 12 G_c\sum_{a=1}^3\sum_{b=0}^3\left[\bar{\psi}_{i,p,\mu}
\psi_{j,q,\nu}\bar{\psi}
_{k,r,\rho}\psi_{l,s,\sigma}\left(\gamma_{\alpha}\right)_{\mu\nu}
\left(\gamma^{\alpha}\right)_{\rho\sigma}\left(t_a\right)_{ij}\left
(t_a\right)_{kl}\left(\tau_b\right)_{pr}\left(\tau_b\right)_{sq}\right]
\eea
In order to Fierz-transform the colour indices we use the relation
\beq
\sum_{a=1}^3\left(t_a\right)_{ij}\left(t_a\right)_{kl}=\frac 12
\left[\delta_{ik}\delta_{lj}+\left(
t_1\right)_{ik}\left(t_1\right)_{lj}+\left(
t_3\right)_{ik}\left(t_3\right)_{lj}\right]-\frac 32
\left(t_2\right)_{ik}\left(t_2\right)_{lj},
\eeq
thus obtaining
\bea
\mathcal{L}_{int}^c&=&-\frac 14 G_c\sum_{b=0}^3\left[\bar{\psi}_{i,p,\mu}
\psi_{j,q,\nu}\bar{\psi}
_{k,r,\rho}\psi_{l,s,\sigma}\left(\gamma_{\alpha}\right)_{\mu\nu}
\left(\gamma^{\alpha}\right)_{\rho\sigma}\left[
\delta_{ik}\delta_{lj}+\right. \right. 
\nonumber\\
&+&\left.\left.\left(
t_1\right)_{ik}\left(t_1\right)_{lj}+\left(
t_3\right)_{ik}\left(t_3\right)_{lj}
-3\left(t_2\right)_{ik}\left(t_2\right)_{lj}\right]
\left(\tau_b\right)_{pr}\left(\tau_b\right)_{sq}\right].
\eea
At the end we perform the Fierz transformation for the Dirac indices and find
\bea
\mathcal{L}_{int}^c&=&-\frac 14 G_c\sum_{b=0}^3\left[\bar{\psi}_{i,p,\mu}
\psi_{j,q,\nu}\bar{\psi}
_{k,r,\rho}\psi_{l,s,\sigma}\left(\left(C^{*}\right)_{\mu\rho}\left(C
\right)_{\sigma\nu}-\frac 12\left(\gamma_{\alpha}C^*\right)_{\mu\rho}
\left(C\gamma^{\alpha}\right)_{\sigma\nu}\right.\right.+
\nonumber\\
&-&\left.\left.\frac 12\left(\gamma_{\alpha}
\gamma_5C^*\right)_{\mu\rho}\left(C\gamma^{\alpha}\gamma_5\right)_{\sigma\nu}
+\left(i\gamma_5C^*\right)_{\mu\rho}\left(iC\gamma_5\right)_{\sigma\nu}\right)
\left(\delta_{ik}\delta_{lj}\right. \right.+
\nonumber\\
&+&\left.\left.\left(
t_1\right)_{ik}\left(t_1\right)_{lj}+\left(
t_3\right)_{ik}\left(t_3\right)_{lj}
-3\left(t_2\right)_{ik}\left(t_2\right)_{lj}\right)
\left(\tau_b\right)_{pr}\left(\tau_b\right)_{sq}\right]=
\nonumber\\
&=&-\frac 14 G_c\sum_{b=0}^3\sum_{S=0,1,3}\left[\left(\bar{\psi}\tau_b
t_SC
\bar{\psi}^T\right)\left(\psi^TC\tau_bt_S\psi\right)+\left(i\bar{\psi}
\gamma_5\tau_bt_SC\bar{\psi}^T\right)\left(i\psi^TC\gamma_5\tau_b
t_S\psi\right)\right.
\nonumber\\
&-&\left.\frac 12\left(\bar{\psi}\gamma_{\alpha}\tau_b
t_SC\bar{\psi}^T\right)\left(\psi^TC\gamma_{\alpha}\tau_b
t_S\psi\right)-\frac 12\left(\bar{\psi}\gamma_{\alpha}\gamma_5\tau_b
t_SC\bar{\psi}^T\right)\left(\psi^TC\gamma_{\alpha}\gamma_5\tau_b
t_S\psi\right)\right]
\nonumber\\
&+&\frac 34 G_c\sum_{b=0}^3\left[
\left(\bar{\psi}\tau_b
t_2C
\bar{\psi}^T\right)\left(\psi^TC\tau_bt_2\psi\right)+\left(i\bar{\psi}
\gamma_5\tau_bt_2C\bar{\psi}^T\right)\left(i\psi^TC\gamma_5\tau_b
t_2\psi\right)\right.
\nonumber\\
&-&\left.\frac 12\left(\bar{\psi}\gamma_{\alpha}\tau_b
t_2C\bar{\psi}^T\right)\left(\psi^TC\gamma_{\alpha}\tau_b
t_2\psi\right)-\frac 12\left(\bar{\psi}\gamma_{\alpha}\gamma_5\tau_b
t_2C\bar{\psi}^T\right)\left(\psi^TC\gamma_{\alpha}\gamma_5\tau_b
t_2\psi\right)\right]
\label{finalfierz1}
\eea
where we have introduced the charge conjugation matrix operator for fermions 
$C=i\gamma_0\gamma_2$.
We can easily read from eq.~(\ref{finalfierz1}) the coefficient of the scalar 
diquark channel:
\beq
H=\frac{3}{2}G_c.
\eeq
Next we show that also $G=3G_c/2$, starting from 
eq.~(\ref{explicit}) and performing a Fierz transformation into the $\bar{q}q$
channel.

We start from the flavour-$SU(2)$ identity
\beq
\delta_{pq}\delta_{rs}=\frac 12\sum_{b=0}^3\left(\tau_b\right)_{ps}
\left(\tau_b\right)_{rq}
\eeq
and obtain
\bea
\mathcal{L}_{int}^c=-\frac 12 G_c\sum_{a=1}^3\sum_{b=0}^3\left[\bar{\psi}_
{i,p,\mu}\psi_{j,q,\nu}\bar{\psi}
_{k,r,\rho}\psi_{l,s,\sigma}\left(\gamma_{\alpha}\right)_{\mu\nu}
\left(\gamma^{\alpha}\right)_{\rho\sigma}\left(t_a\right)_{ij}\left
(t_a\right)_{kl}\left(\tau_b\right)_{ps}\left(\tau_b\right)_{rq}\right].
\eea
Then we transform colour indices by using
\beq
\sum_{a=1}^3\left(t_a\right)_{ij}\left(t_a\right)_{kl}=
\frac{3}{2}\delta_{il}\delta_{kj}-\frac 12\sum_{c=1}^3\left(t_c
\right)_{il}\left(t_c\right)_{kj}
\eeq
and find
\bea
\mathcal{L}_{int}^c&=&-\frac 14 G_c\sum_{b=0}^3\left[\bar{\psi}_{i,p,\mu}
\psi_{j,q,\nu}\bar{\psi}
_{k,r,\rho}\psi_{l,s,\sigma}\left(\gamma_{\alpha}\right)_{\mu\nu}
\left(\gamma^{\alpha}\right)_{\rho\sigma}\left(3\delta_{il}
\delta_{kj}\right.\right.
\nonumber\\
&-&\left.\left.\sum_{c=1}^3\left(t_c\right)_{il}
\left(t_c\right)_{kj}\right)\left(\tau_b\right)_{ps}\left(\tau_b
\right)_{rq}\right].
\eea
Finally the Dirac Fierz transformation leads to:
\bea
\mathcal{L}_{int}^c&=&-\frac 14 G_c\sum_{b=0}^3\left[\bar{\psi}_{i,p,\mu}
\psi_{j,q,\nu}\bar{\psi}
_{k,r,\rho}\psi_{l,s,\sigma}\left(\delta_{\mu\sigma}\delta_{\rho\nu}-\frac 12
\left(\gamma_{\alpha}\right)_{\mu\sigma}\left(\gamma^{\alpha}\right)_{\rho\nu}-
\frac 12\left(\gamma_{\alpha}\gamma_5\right)_{\mu\sigma}\left(\gamma^{\alpha}
\gamma_5\right)_{\rho\nu}+\right.\right.
\nonumber\\
&+&\left.\left.\left(i\gamma_5\right)_{\mu\sigma}\left(i
\gamma_5\right)_{\rho\nu}\right)\left(3\delta_{il}\delta_{kj}
-\sum_{c=1}^3\left(t_c\right)_{il}
\left(t_c\right)_{kj}\right)\left(\tau_b\right)_{ps}\left(\tau_b
\right)_{rq}\right]=
\nonumber\\
&=&\frac 14 G_c\sum_{b=0}^3\left\{3\left[\left(\bar{\psi}\tau_b\psi\right)^2+
\left(i\bar{\psi}\gamma_5\tau_b\psi\right)^2-\frac 12
\left(\bar{\psi}\gamma_{\alpha}\tau_b\psi\right)^2-\frac 12\left(\bar{\psi}
\gamma_{\alpha}\gamma_5\tau_b\psi\right)^2\right]\right.
\nonumber\\
&-&\left.\sum_{a=1}^3
\left[\left(\bar{\psi}t_a\tau_b\psi\right)^2+\left(i
\bar{\psi}\gamma_5t_a\tau_b\psi\right)^2-\frac 12\left(\bar{\psi}
\gamma_{\mu}t_a\tau_b\psi\right)^2-\frac 12\left(\bar{\psi}\gamma_{\mu}
\gamma_5t_a\tau_b\psi\right)^2\right]\right\}
\nonumber
\eea
from which we can easily read 
\beq
G=\frac 32 G_c.
\eeq
\bibliography{biblio}
\bibliographystyle{h-physrev3}
\end{document}